# $Cu_{1-x}Al_x$ films as Alternatives to Copper for Advanced Interconnect Metallization

Jean-Philippe Soulié [1,a)], Kiroubanand Sankaran [1], Geoffrey Pourtois [1], Johan Swerts [1], Zsolt Tőkei [1], and Christoph Adelmann [1]

[1] Imec, 3001 Leuven, Belgium

a) Electronic mail: jean-philippe.soulie@imec.be

$Cu_xAl_{1-x}$ thin films with $0.2 \leq x \leq 0.7$ have been studied as potential alternatives for the metallization of advanced interconnects. First-principles simulations were used to obtain the $Cu_xAl_{1-x}$ electronic structure and cohesive energy to benchmark different intermetallics and their prospects for interconnect metallization. Next, thin $Cu_xAl_{1-x}$ films were deposited by PVD with thicknesses in the range between 3 and 28 nm. The lowest resistivities of 9.5 µΩcm were obtained for 28 nm thick stochiometric CuAl and $CuAl_2$ after 400°C post-deposition annealing. Based on the experimental results, we discuss the main challenges for the studied aluminides from an interconnect point of view, namely the control of the film stoichiometry, the phase separation observed for off-stoichiometric CuAl and $CuAl_2$, as well as the presence of a nonstoichiometric surface oxide.

# I. INTRODUCTION

In the past few decades, Cu has been the conventional choice for interconnect metallization. However, as interconnect dimensions are continuously reduced, the Cu line resistance increases rapidly due to scaling limits of conductive barrier and liner layers, as well as the strong size effect of the Cu resistivity. This has a large impact on microelectronic circuit performance, leading to higher latency and elevated heat dissipation. Moreover, Cu shows reduced electromigration performance at nanoscale dimensions, limiting the reliability of Cu-based interconnects.[1-3] The pursuit of alternative metallization schemes thus received increasing attention as state-of-the-art interconnects in advanced technology reach sub-10 nm dimensions, where barrier and liner scaling reaches its limit.[4] Initial investigations have focused on elemental metals with short mean free paths (MFPs) of the charge carriers, expected to show reduced finite size effects on the resistivity, and elevated melting temperatures.[5-8] Notably, metals such as Co,[9] Pt-group metals, particularly Ru,[10] and Mo,[11] can indeed exhibit superior properties compared to Cu at small dimensions. However, Co still necessitates the incorporation of a diffusion barrier layer, while Ru mandates the use of a liner layer for optimal adhesion.[12] More recently, research efforts have transitioned towards the exploration of binary and ternary compounds. Within compound metals, only ordered intermetallics possess, however, low bulk resistivities and are thus potentially attractive for interconnect metallization applications. Both stoichiometric and nonstoichiometric disorder can induce alloy scattering, leading to a significant rise in resistivity.[13]

In particular aluminide intermetallics, such as NiAl, ScAl$_3$, CuAl, and CuAl$_2$ have recently garnered considerable attention.[14-16] Here, we explore the Cu-Al system for potential applications in interconnects. Thin films of CuAl$_2$ have shown low resistivity, high resistance to electromigration (EM), and good reliability against bias-temperature stress (BTS).[17,18] Its diffusion barrier properties are attributed to the formation of a dense and self-limited Al oxide layer at the CuAl$_2$/SiO$_2$ interface.[18] In addition, CuAl has also been proposed as a potential candidate for interconnect metallization,[19] with good adhesion to oxides, high BTS reliability, and promising EM resistance. For these reasons, both stochiometric CuAl$_2$ and CuAl emerge as potential candidates for a liner-free intermetallic compound in future technology nodes.

Besides CuAl$_2$ and CuAl, the Cu-Al phase diagram [20] contains several other intermetallic phases in the temperature range between 400°C and 500ºC, relevant for logic interconnect applications. We have therefore performed an *ab initio* screening exercise to identify the most promising compounds. This is then followed by a thin film exploration of the properties of Cu$_{1-x}$Al$_x$ (0.2 ≤ x ≤ 0.7) compounds at thicknesses in the range between 3 and 28 nm. We show that, rather common to many aluminides, several main challenges remain to control the stoichiometry and phase of the films, such as nonstoichiometric surface oxidation and interfacial reactions with underlying dielectrics, but also phase separation as both CuAl$_2$ and CuAl have very narrow phase fields.

## II. EXPERIMENTAL AND MODELING DETAILS

All $Cu_{1-x}Al_x$ films were deposited onto 300 mm Si (100) wafers using a Canon Anelva physical-vapor deposition (PVD) system at room temperature by co-sputtering from Al and Cu targets. Prior to $Cu_{1-x}Al_x$, a 100 nm thick thermal $SiO_2$ layer was grown on the Si wafers to ensure electrical insulation. Fluxed were calibrated using Rutherford backscattering spectrometry. Post-deposition annealing was carried out in $H_2$ for 30 minutes at atmospheric pressure. Wavelength-dispersive x-ray analysis (WDX) confirmed compositions deduced from flux calibrations with an absolute accuracy better than 0.5% Al and a precision (repeatability) better than 0.1%. Crystal structure and film thicknesses were determined by grazing-incidence x-ray diffraction (GIXRD, $\omega = 0.3°$) and x-ray reflectance (XRR), respectively, utilizing a Bruker JVX7300 diffractometer with Cu K$\alpha$ radiation. Film thickness measurements using XRR before and after annealing showed no variation within experimental accuracy. Further insight into crystallinity, thickness, and chemical composition were obtained through transmission electron microscopy (TEM) using a FEI Titan electron microscope at 200 kV. The film resistivity was derived from sheet resistance (Rs) measurements using a KLA Tencor RS100 system, combined with XRR film thickness, corrected for the thickness of the surface oxide, which was assumed to be insulating.

Electronic band structures were computed using density functional theory (DFT) simulations, as implemented in QUANTUM ESPRESSO DFT package [21] with a kinetic cutoff energy ranging between 60 and 80 Ry for the truncation of the plane-wave expansion of the wavefunction including a Methfessel-Paxton smearing function with a broadening of 13.6 meV. The valence electrons were represented by Garrity-Bennet-Rabe-Vanderbilt pseudopotentials [22] together with the exchange-correlation energy described within the Perdew-Burke-Ernzerhof generalized gradient approximation.[23] The first Brillouin zone was sampled using a regular and unshifted Monkhorst-Pack scheme [24] with a *k*-point density ranging from 25×25×25 to 61×61×61.

Pre-optimized primitive crystal structures of $Cu_{1-x}Al_x$ intermetallics were obtained from the Materials Project online database.[25] Their electronic properties were then computed using automated DFT simulations to evaluate the $\rho_0 \times \lambda$ tensor for potential resistivity scalability. In addition, the cohesive energy was computed, which has been used as a proxy for EM resistance.

## III. RESULTS AND DISCUSSION

The scaling potential of various $Cu_{1-x}Al_x$ intermetallics in the studied composition range has been assessed using the methodology introduced by D. Gall [26] and extended by K. Moors *et al.* [27] for different transport directions within a finite temperature range set around room temperature (300 K). The potential for resistivity scalability can be represented by the components of the $\rho_0 \times \lambda$ transport tensor, where $\rho_0$ is the bulk resistivity and $\lambda$ the electron mean-free path. The tensor form naturally considers both the symmetry group of the material and the anisotropy of the conductivity. We have evaluated the generalized $\rho_0 \times \lambda$ tensor within the constant $\lambda$ approximation, considering arbitrary transport directions,[27] which has the general symmetric form:

$$\underline{\underline{\rho_0 \times \lambda}} = \begin{pmatrix} (\rho_0\lambda)_{xx} & (\rho_0\lambda)_{xy} & (\rho_0\lambda)_{xz} \\ (\rho_0\lambda)_{xy} & (\rho_0\lambda)_{yy} & (\rho_0\lambda)_{yz} \\ (\rho_0\lambda)_{xz} & (\rho_0\lambda)_{yz} & (\rho_0\lambda)_{zz} \end{pmatrix} \qquad (1)$$

Figure 1 summarizes the computed diagonal components of the $\rho_0 \times \lambda$ tensors of different $Cu_{1-x}Al_x$ intermetallic phases together with their cohesive energy. Full $\rho_0 \times \lambda$ tensors reported in Table 1. Several cubic and tetragonal phases, namely $CuAl_2$, $Cu_3Al$, and $CuAl_3$ show diagonal components of the $\rho_0 \times \lambda$ tensors with lower (better) values than Cu (dashed black line). Hence, these compounds promise good resistivity scalability in narrow dimensions. By contrast, CuAl as well as $Cu_3Al$ and $Cu_2Al_3$ for some transport directions, are less promising since the $\rho_0 \times \lambda$ value exceeds that of Cu. In addition, all $Cu_{1-x}Al_x$ alloys possess similar cohesive energies as Cu (blue line), which suggests that the EM performance of these intermetallic phases needs to be carefully considered. Nonetheless, first experiments have shown promising EM performance for CuAl and $CuAl_2$.[17,19]

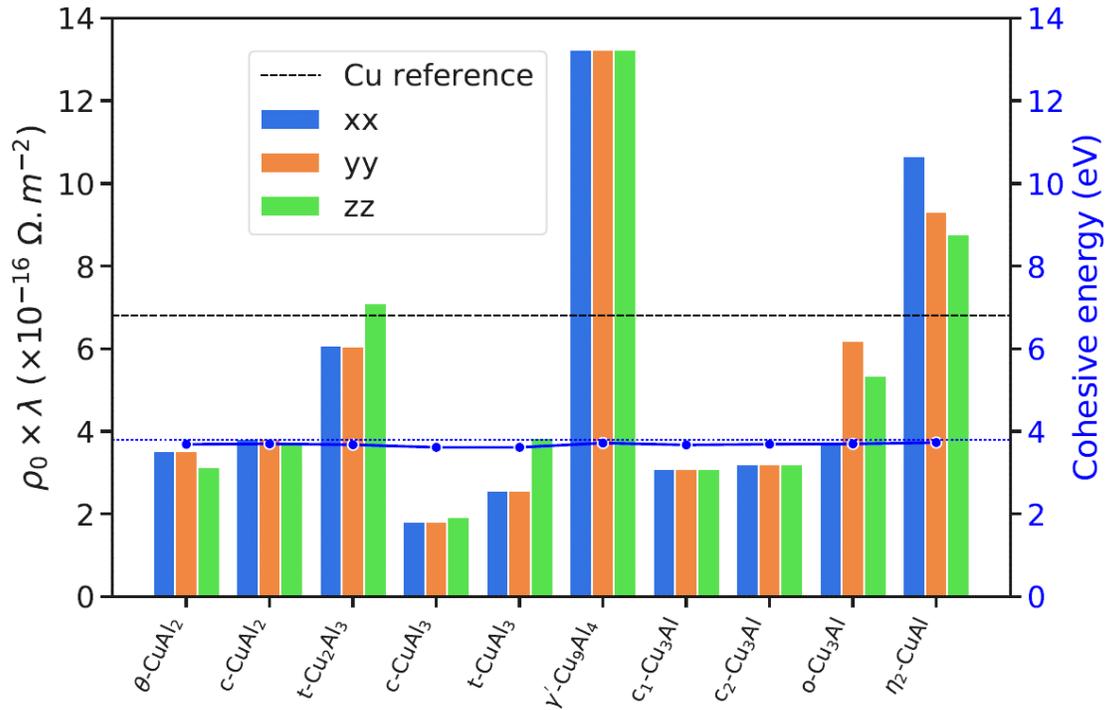

Fig. 1. Computed diagonal components of $\rho_0 \times \lambda$ tensors and cohesive energies of different stoichiometries and symmetries as cubic (c), tetragonal (t), orthorhombic (o), and phase

name for experimentally known (θ, γ, η) $Cu_{1-x}Al_x$ alloys together with Cu (dashed lines) as reference for direct comparison.

| Compound | Materials Project ID | Symmetry (group) | $\rho_0 \times \lambda$ ($\times 10^{-16}$ $\Omega.m^{-2}$) | Cohesive energy (eV) |
|---|---|---|---|---|
| c-$CuAl_2$ | mp-985806 | Cubic (225) | $\begin{pmatrix} 3.8 & 0 & 0 \\ 0 & 3.8 & 0 \\ 0 & 0 & 3.8 \end{pmatrix}$ | 3.7 |
| γ'-$Cu_9Al_4$ | mp-553 | Cubic (215) | $\begin{pmatrix} 13.2 & 0 & 0 \\ 0 & 13.2 & 0 \\ 0 & 0 & 13.2 \end{pmatrix}$ | 3.7 |
| $c_1$-$Cu_3Al$ | mp-12777 | Cubic (225) | $\begin{pmatrix} 3.1 & 0 & 0 \\ 0 & 3.1 & 0 \\ 0 & 0 & 3.1 \end{pmatrix}$ | 3.6 |
| $c_2$-$Cu_3Al$ | mp-1008555 | Cubic (221) | $\begin{pmatrix} 3.2 & 0 & 0 \\ 0 & 3.2 & 0 \\ 0 & 0 & 3.2 \end{pmatrix}$ | 3.7 |
| θ-$CuAl_2$ | mp-998 | Tetragonal (140) | $\begin{pmatrix} 3.5 & 0 & 0 \\ 0 & 3.5 & 0 \\ 0 & 0 & 3.1 \end{pmatrix}$ | 3.7 |
| $t_1$-$CuAl_3$ | mp-1183161 | Tetragonal (123) | $\begin{pmatrix} 1.7 & 0 & 0 \\ 0 & 1.7 & 0 \\ 0 & 0 & 1.9 \end{pmatrix}$ | 3.6 |
| $t_2$-CuAl$_3$ | mp-985825 | Tetragonal (123) | $\begin{pmatrix} 2.6 & 0 & 0 \\ 0 & 2.6 & 0 \\ 0 & 0 & 3.8 \end{pmatrix}$ | 3.6 |
| t- Cu$_2$Al$_3$ | mp-10886 | Trigonal (164) | $\begin{pmatrix} 6.0 & 0 & 0 \\ 0 & 6.0 & 0 \\ 0 & 0 & 7.2 \end{pmatrix}$ | 3.7 |
| o-Cu$_3$Al | mp-12802 | Orthorhombic (59) | $\begin{pmatrix} 3.7 & 0 & 0 \\ 0 & 6.2 & 0 \\ 0 & 0 & 5.3 \end{pmatrix}$ | 3.7 |
| $\eta_2$-CuAl | mp-2500 | Monoclinic (23) | $\begin{pmatrix} 10.7 & 0 & 51.6 \\ 0 & 9.3 & 0 \\ 51.6 & 0 & 8.8 \end{pmatrix}$ | 3.7 |

Table 1. Detailed components of $\rho 0 \times \lambda$ tensors and cohesive energies of different stoichiometries and symmetries as cubic (c), tetragonal (t), orthorhombic (o), and phase name for experimentally known ($\theta$, $\gamma$, $\eta$) Cu$_{1-x}$Al$_x$ alloys.

Before discussing thin film resistivities, we address the phase formation in the studied composition range. Figure 2a shows GIXRD pattern of 28 nm thick Cu$_{1-x}$Al$_x$ films with Al mole fractions $x \approx 0.25$ (around stoichiometric Cu$_3$Al). As deposited, patterns are nearly x-ray amorphous with only one weak peak visible around 44°. After 500°C post-deposition annealing in H$_2$, the peak intensified for compositions of $x = 0.25$ and above, while films with less Al remained x-ray amorphous (Fig. 2b). The peak can be attributed to the Cu$_3$Al (220) reflection, although it cannot be ruled out that it stems from the cubic Cu$_9$Al$_4$ phase ((330) reflection), which may have formed due to phase separation.[20]

Similarly, as-deposited $Cu_{1-x}Al_x$ thin films with compositions around stochiometric CuAl ($x \approx 0.50$) showed poor crystallinity but crystallized after annealing at 500°C in $H_2$ (Figs. 3a and 3b). The GIXRD patterns after annealing were consistent with the expected $\eta_2$-CuAl phase. By contrast, films with compositions around $CuAl_2$ ($x \approx 0.66$) were already polycrystalline as deposited (Fig. 4a). Post-deposition annealing at 500°C in $H_2$ further improved the crystallinity of the films without changing the phase (Fig. 4b). Also in this case, the GIXRD patterns were consistent with the observation of the $\theta$-$CuAl_2$ phase. Interestingly, at compositions between stoichiometric CuAl and $CuAl_2$, the GIXRD patterns indicated a two-phase region with both CuAl and $CuAl_2$ present. This is in agreement with the Cu-Al phase diagram,[20] which indicates narrow single phase regions for CuAl and $CuAl_2$, which are essentially line compounds. We note that the high-temperature orthorhombic $\eta_1$-CuAl phase was not visible in the patterns, although it is difficult to completely rule out its presence due to peak overlaps.

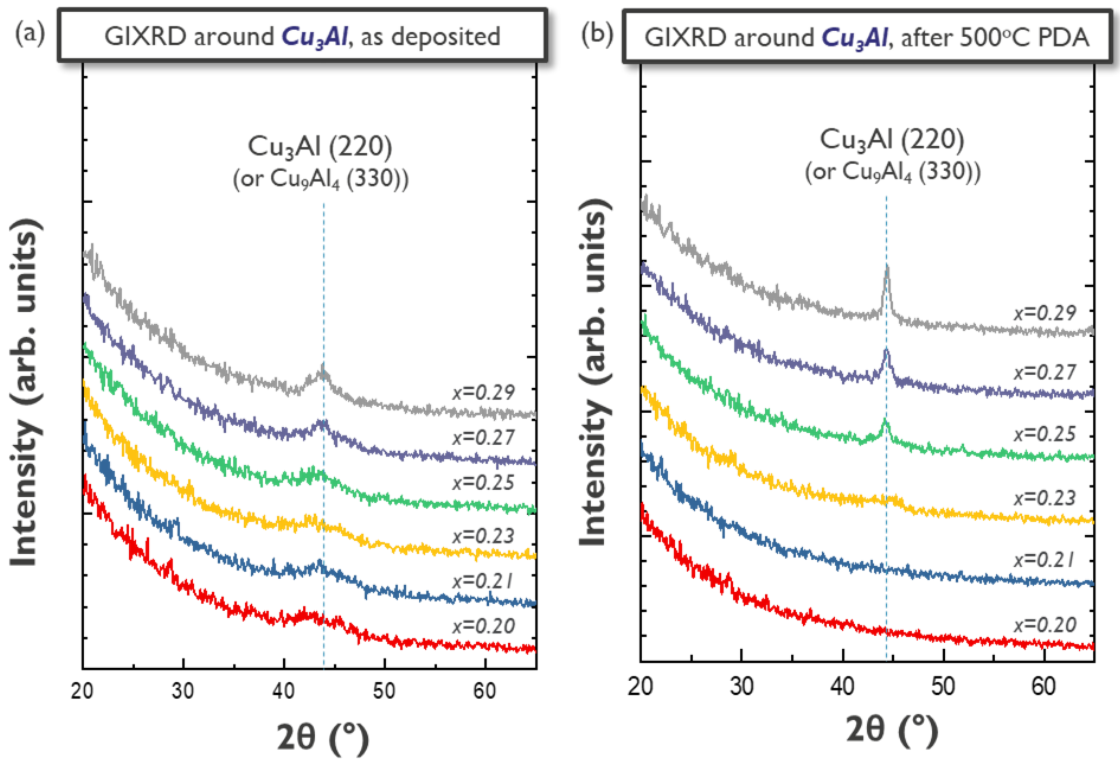

Fig. 2. GIXRD patterns of 28 nm thick $Cu_{1-x}Al_x$ (0.20 ≤ x ≤ 0.29) films as a function of Al mole fraction *x* (a) as deposited and (b) after post-deposition annealing at 500°C in $H_2$.

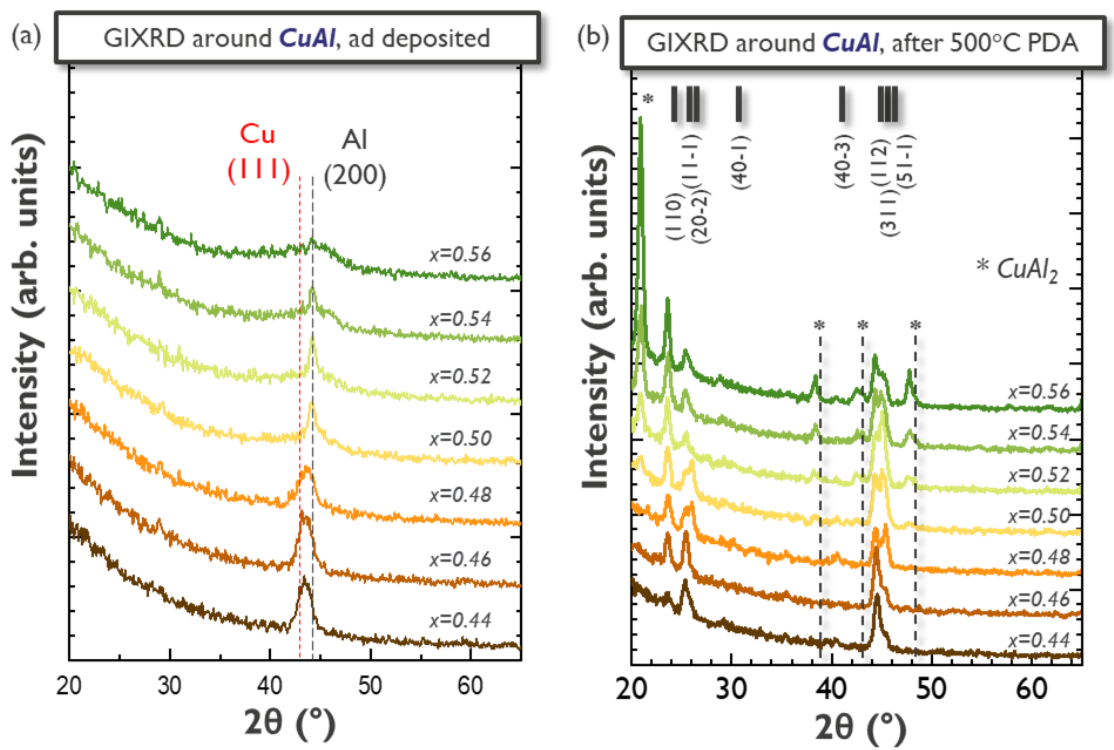

Fig. 3. GIXRD patterns of 28 nm thick $Cu_{1-x}Al_x$ ($0.44 \leq x \leq 0.56$) films as a function of Al mole fraction $x$ (a) as deposited and (b) after post-deposition annealing at 500°C in $H_2$.

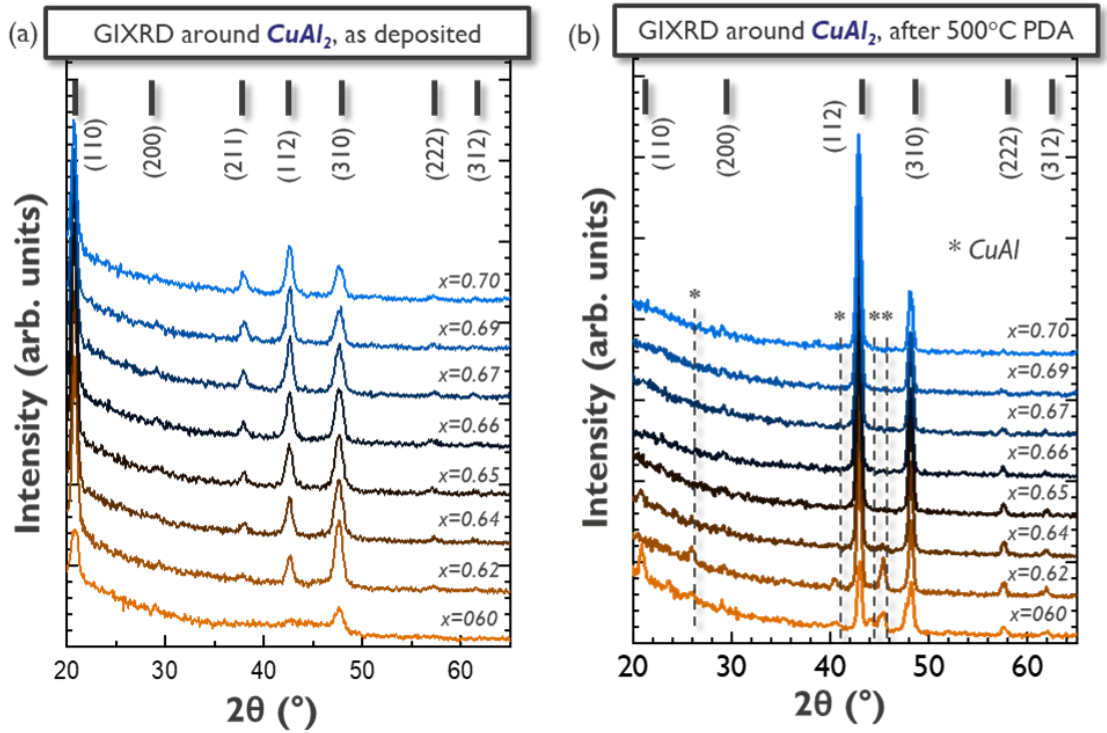

Fig. 4. GIXRD patterns of 28 nm thick $Cu_{1-x}Al_x$ ($0.60 \leq x \leq 0.70$) films as a function of Al mole fraction $x$ (a) as deposited and (b) after post-deposition annealing at 500°C in $H_2$.

Figure 5 shows the resistivity of 28 nm thick $Cu_{1-x}Al_x$ films with Al mole fractions $x$ between 0.2 and 0.7, both as deposited at room temperature and after different post-deposition annealing temperatures. Resistivities generally decreased with annealing temperature because of crystallization and grain growth, as well as possibly ordering and defect annihilation.[28] Around $Cu_3Al$ the resistivity showed a sharp reduction of the resistivity after annealing at 400°C with some gradual further reduction up to 600°C. At the highest annealing temperature, a narrow resistivity minimum became visible around $x$ = 0.25, which is indicated of the formation of an ordered intermetallic. The lowest observed resistivity was 14.0 µΩcm after 600°C post-deposition annealing (28 nm film thickness).

A large decrease of the resistivity was also observed for compositions around stochiometric CuAl films between room temperature and after annealing at 400°C. Annealing at 500°C showed only a minor resistivity reduction, indicating a saturation effect for the annealing temperature. By contrast, the resistivity of as deposited $CuAl_2$ was already low around stoichiometry ($x$ = 0.66), which is consistent with the observation of a polycrystalline microstructure already as deposited. The narrow minimum around $x$ = 0.66 is again indicative of the formation of an ordered intermetallic. The lowest resistivity in the studied composition range was observed to be 9.5 µΩcm after annealing at 500°C in the region around CuAl and $CuAl_2$, remarkably rather independent of composition. We note that the observed resistivity for films as thin as 28 nm was very close to the bulk resistivity of CuAl (8.7 µΩcm),[29] while still comparable to that of $CuAl_2$ (6.5 µΩcm).[30] Recently, some resistivity values were reported for thicker film than our study: for a 100nm of $CuAl_2$ film after annealing at 400°C, resistivity was equal to 7.8 µΩcm [17, 31] and 12.2 µΩcm for 100nm of CuAl annealed in the same conditions.[19]

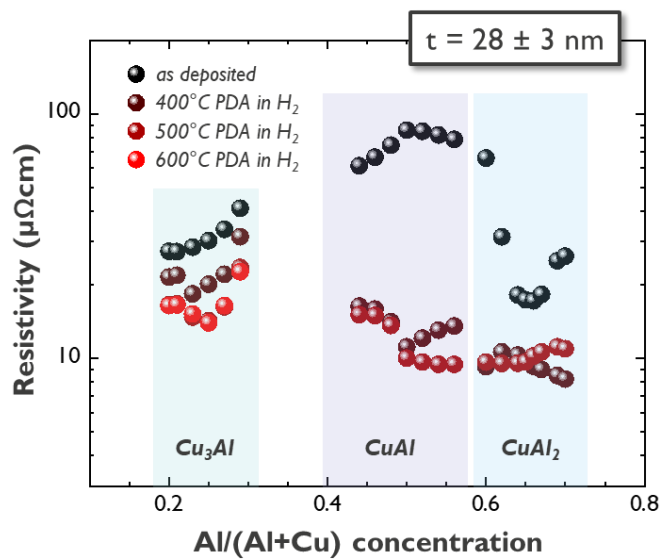

Fig. 5. $Cu_{1-x}Al_x$ resistivity *vs.* Al mole fraction *x* for 28 nm thick films both as deposited and after annealing in $H_2$ at indicated temperatures.

The microstructure and the chemical profile of the 28 nm thick $CuAl_2$ and CuAl films were further investigated by cross-sectional TEM imaging and EDS (Fig. 6). The TEM images of $CuAl_2$ after 500°C anneal in Fig. 6a show large grains with bamboo-like grain boundaries, of the same size than the thickness of the film. Chemical profiling using electron-dispersive spectroscopy (EDS), in Fig. 6b, indicated that the central part of the film is uniformly consistent with the stoichiometric ratio Al/Cu = 2 within experimental accuracy. By contrast, the high-angle annular dark field scanning TEM (HAADF-STEM) image for CuAl showed a strong contrast change within the image indicating a chemical composition variation (Fig. 6c). This can be attributed to two grains with different compositions. Analyzing both grains with EDS found that one grain has an Al/Cu ratio of 1 within experimental accuracy, whereas the other grain has a ratio of 2:1. This can be attributed to a secondary $CuAl_2$ minority in CuAl, in agreement with the GIXRD patterns in Fig. 3b, possibly due to small deviations from stoichiometry. This is in agreement with the phase diagram [20] since both CuAl and $CuAl_2$ have very narrow phase regions. Any intermediate stoichiometry should thus lead to phase separation and formation of a $CuAl/CuAl_2$ mixed phase. This can explain the observation that the resistivity in the composition range between CuAl and $CuAl_2$ is rather insensitive to composition. In this range, films contain a mixture of stoichiometric CuAl and $CuAl_2$ films with varying relative volume, as given by the lever rule. Since CuAl and $CuAl_2$ possess rather similar resistivities in thin film form, the mixed phase resistivity is characterized by a weak composition dependence of the resistivity. We note however, that recently, the effect of composition deviation of $CuAl_2$ has been studied on electromigration.[32] The study found that Cu-rich compositions enhance electromigration properties. Therefore, the composition of CuAl and $CuAl_2$ can be expected to have significant effect on electromigration, in contrast to the resistivity.

The TEM images (Figs. 6a and 6d) also showed the presence of an about 4 nm thick native oxide on the surface of the stack, and an about 1-2 nm thick interfacial layer in contact with the underlying $SiO_2$. This interfacial layer leads to very strong adhesion of $CuAl_2$ on $SiO_2$, both as deposited and after 400°C PDA. Measurements by 4-point probe bending found an adhesion energy of 17 $J/m^2$ as deposited and 20 $J/m^2$ after annealing. The interfacial layer is mostly formed by alumina with a small concentration of Si. Since it can act as a passivation layer, $CuAl_2$ is not expected require further liner or barrier layers, as reported before,[17] with excellent time-dependent dielectric breakdown (TDDB) reliability.[33] It should however note that the interfacial layer is not expected to contribute to the interconnect conductance and its thickness must therefore strongly minimized.

EDS further shows that the native oxide on top of the layer after exposure to air contains only alumina with Cu incorporation below detection limits. This oxide layer has been studied in more details by K. Son *et al.* [34] although the studied film was much thicker (382 nm) than in our study. We note that the formation of the nonstoichiometric surface oxide can be attributed to element-specific diffusion and reaction kinetics, similar to what has been observed for NiAl [35] and $Al_3Sc$ [36]. While the surface oxide again is expected to passivate the film, its nonstoichiometric composition renders the control of the film composition more difficult as oxidation (*e.g.* in air) simultaneously leads to Al loss of the bulk of the film. Hence, for microelectronic applications, surface oxidation should be avoided.

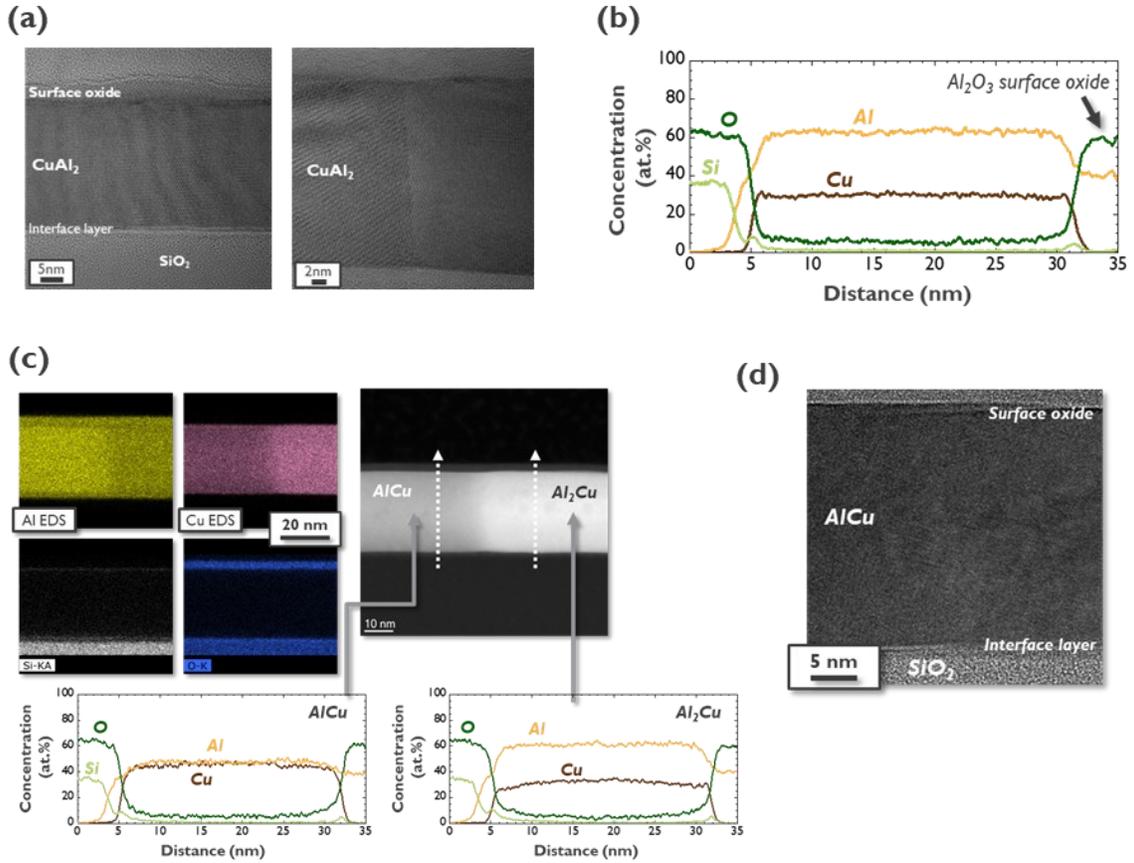

Fig. 6. (a) Cross-sectional TEM images of a CuAl$_2$ film after annealing at 500°C. (b) Composition profile across the CuAl$_2$ film obtained by energy-dispersive spectroscopy (EDS). (c) High-angle annular dark field scanning TEM (HAADF-STEM) image of the CuAl film with EDS related to each contrast. (d) Cross-sectional TEM image of a CuAl film.

Finally, to further assess the potential of $Cu_{1-x}Al_x$ for scaled interconnects, the resistivity was determined at thicknesses down to 4 nm (Fig. 7). Two regimes were observed, similar to previously reported results for NiAl [35] and ScAl$_3$ [36], with a strong increase of the resistivity below 10 nm thickness. This can be understood by the formation of a nonstoichiometric surface oxide as well as an Al-rich interface layer (see Fig. 6), which can alter the composition of the film strongly due to Al loss. Nonetheless, CuAl and $CuAl_2$ films with thicknesses above 10 nm show resistivities below 20 μΩcm and reach values of less than 10 μΩcm around 30 nm after 500°C annealing. We remark that the $CuAl_2$ resistivity outperformed Ru at 10 nm film thickness and above, while CuAl and $CuAl_2$ outperformed Mo for the whole studied thicknesses range. Both compounds also show a similar resistivity as TaN/Cu/TaN below about 8 nm. One has to note that the films appeared x-ray amorphous below 10 nm (not shown here) although it is difficult to conclude unambiguously due to the small diffracting volume. The loss of crystallinity can contribute to a rapid resistivity increase as seen on the mentioned trend, as well as lack of stoichiometry control in ultrathin films because of Al loss to the surface oxide, as observed already in the NiAl system.[35] This demonstrates that the stoichiometric composition control (including the phase separation seen in the $Cu_xAl_{1-x}$ system) and the optimization of crystallinity, grain structure, and ordering can have large effects on the resistivity and are therefore the main challenges to obtain low resistivity $Cu_{1-x}Al_x$ intermetallics and transition metal aluminides in general.

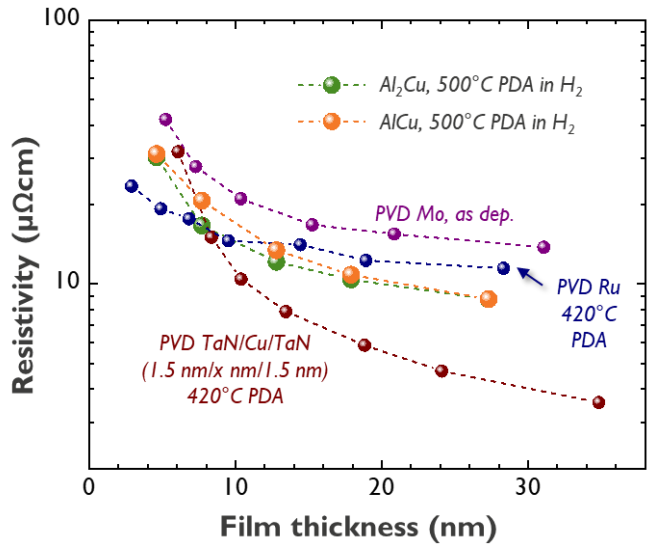

Fig. 7. Resistivity of CuAl and CuAl$_2$ thin films vs. film thickness after PDA at 500°C in H$_2$ compared with elemental metals (TaN/Cu/TaN and Ru data after 420°C PDA and Mo as deposited at room temperature.

## IV. SUMMARY AND CONCLUSIONS

In conclusion, we have reported the properties of Cu$_x$Al$_{1-x}$ thin films with $0.2 \leq x \leq 0.7$ deposited by PVD as a potential alternatives for the metallization of interconnects. The lowest resistivities obtained for 28 nm thick stochiometric CuAl and CuAl$_2$ were around 9.5 μΩ cm after 400°C post-deposition annealing in H$_2$. The resistivity of CuAl$_2$ was below that of Ru thin films for thicknesses above 10 nm, while both CuAl and CuAl$_2$ outperformed Mo for the whole range of thicknesses examined. Moreover, both Cu aluminides showed lower resistivity than TaN/Cu/TaN stacks below about 8 nm film thickness.

Yet, similar to other aluminides,[35,36] certain challenges arise for successful integration of such materials into interconnects. Main challenges include controlling the stoichiometry of ultrathin films, limiting surface oxidation, and managing phase separation. These hurdles can potentially be mitigated by implementing *in situ* capping layers and/or by reducing the required thermal budget to achieve low resistivity films. Employing such solutions will ultimately enable the evaluation of the intrinsic scaling potential of the resistivity and the reliability of these intermetallics in interconnect lines. We finally note that also the two-phase region between CuAl and $CuAl_2$ is of potential interest to applications since the resistivity is nearly independent of composition, strongly facilitating the optimization of the resistivity within-wafer and wafer-to-wafer uniformity. However, dealing with two distinct phases also leads to many challenges for unit process steps, such as reactive-ion etching or surface cleaning, which may have different reaction rates for the two phases. Future process development will therefore be needed to establish the potential of such two-phase regions for practical interconnect applications.

## ACKNOWLEDGMENTS

This work was supported by imec's industrial affiliate program on nano-interconnects. The authors would like to thank Maxim Korytov for the TEM analysis; Kris Vanstreels and Myriam Van De Peer for the adhesion test; and the imec pilot line for their support.

## AUTHOR DECLARATIONS

**Conflicts of Interest** *(required)*

The authors have no conflicts to disclose.

# DATA AVAILABILITY

The data that support the findings of this study are available from the corresponding author upon reasonable request.